\documentclass[
]{ceurart}

\usepackage{subfigure}
\usepackage{soul}
\usepackage{tabularx}
\begin{document}

\copyrightyear{2024}
\copyrightclause{Copyright for this paper by its authors.
  Use permitted under Creative Commons License Attribution 4.0
  International (CC BY 4.0).}

\conference{NORMalize 2024: The Second Workshop on Normative Design and Evaluation of Recommender Systems, October 18th, 2024, co-located with the ACM Conference on Recommender Systems 2024 (RecSys 2024), Italy}

%\title{An Opensource Library for Generating Synthetic Sparse Categorical Datasets}
\title{Generating Diverse Synthetic Datasets for Evaluation of Real-life Recommender Systems}
\author[1]{Miha Malenšek}[%
orcid=0009-0004-6941-2203,
email=mihamalen@gmail.com
]
\cormark[1]
\address[1]{Faculty of Computer and Information Science, Ljubljana, Slovenia}
\author[2]{Blaž Škrlj}[%
orcid=0000-0002-9916-8756,
email=bskrlj@outbrain.com
]
\author[2]{Blaž Mramor}[%
orcid=0009-0001-9669-5374,
email=blazmramor@hotmail.com
]
\address[2]{Outbrain, Ljubljana, Slovenia}
\author[1]{Jure Demšar}[%
orcid=0000-0001-6080-149X,
email=jure.demsar@fri.uni-lj.si
]

\begin{abstract}
    Synthetic datasets are important for evaluating and testing machine learning models. When evaluating real-life recommender systems, high-dimensional categorical (and sparse) datasets are often considered. Unfortunately, there are not many solutions that would allow generation of artificial datasets with such characteristics. For that purpose, we developed a novel framework for generating synthetic datasets that are diverse and statistically coherent. Our framework allows for creation of datasets with controlled attributes, enabling iterative modifications to fit specific experimental needs, such as introducing complex feature interactions, feature cardinality, or specific distributions. We demonstrate the framework's utility through use cases such as benchmarking probabilistic counting algorithms, detecting algorithmic bias, and simulating AutoML searches. Unlike existing methods that either focus narrowly on specific dataset structures, or prioritize (private) data synthesis through real data, our approach provides a modular means to quickly generating completely synthetic datasets we can tailor to diverse experimental requirements. Our results show that the framework effectively isolates model behavior in unique situations and highlights its potential for significant advancements in the evaluation and development of recommender systems. The readily-available framework is available as a free open Python package to facilitate research with minimal friction.
\end{abstract}


\maketitle

\section{Introduction}
\label{sec:intro}

Recommender systems have significantly enhanced user experiences across various digital platforms, from e-commerce to streaming services and advertisement. These systems analyze user preferences and behaviours to suggest relevant items, thereby increasing user engagement and satisfaction. However, evaluating and benchmarking these systems poses unique challenges due to the sensitivity and availability of real-world data. Privacy concerns, data access restrictions, and the need for diverse testing scenarios often limit researchers' ability to conduct comprehensive evaluations and perform thorough benchmarking.

Synthetic data generation offers a promising solution to these challenges and is already widely used in fields like computer vision and robotics \cite{lesnikowski2021synthetic}, yet still in its nascent stages in the field of recommender systems. Prior research in the field mainly focused on maintaining data fidelity and ensuring user privacy. For instance, Slokom et al. \cite{slokom2018comparing} introduced the SynRec framework to generate partially synthetic data using CART, while Berlioz et al. \cite{berlioz2015applying} applied differential privacy techniques to protect user information with matrix factorization. Efforts to scale academic datasets to production standards include Antulov-Falin et al.'s \cite{antulov2014synthetic} memory-biased random walks and Belletti et al.'s \cite{belletti2019scalable} fractal expansions. Provalov et al. \cite{provalov2021synevarec} developed the SynEva framework using GAN techniques for synthetic data generation based on real data. All this successful research point to significant potential for advancements in recommender systems through synthetic data, warranting further research on data fidelity, privacy-preserving techniques, and new simulation methods.

Despite this, there remains a need for frameworks that generate completely synthetic data tailored specifically to the evaluation of recommender systems. Such data should possess characteristics that make it suitable for rigorous testing and algorithmic development, without relying on privacy-preserving transformations of real data. This paper addresses this gap by
proposing a comprehensive framework for generating diverse and statistically coherent synthetic datasets tailored to the evaluation of recommender systems. Unlike techniques that focus on privacy or emulate real data, our approach creates entirely artificial, production-scale data with specific qualities and attributes. Controlling the generation process allows iterative modifications to fit specific experimental needs, such as introducing complex feature interactions or increasing the cardinality of the dataset. Our deterministic generative process allows for reproducibility and enables on-the-fly dataset modifications, reducing setup time for experimental scenarios. We demonstrate the framework's utility through use cases in benchmarking algorithms, detecting algorithmic bias, and simulating AutoML searches.

\section{Generating Categorical Datasets}
\label{sec:gener}

To streamline the creation of synthetic datasets for classification tasks and avoid reliance on imperfect or complex real-world data samples, we introduce a comprehensive framework called \textit{CategoricalClassification}. The framework is available through the Python Package Index (PyPI, \url{https://pypi.org/}) and can be installed with \textit{pip}. The core of our solution is packaged into a Python class called \textit{CategoricalClassification}. Functionalities implemented in this class allow for rapid generation of production-scale synthetic datasets with specific attributes tailored to the nature of the problem, such as sparsity, high cardinality features or specific distributions. Additionally, our framework offers functionalities to augment these datasets by incorporating feature combinations, correlations, and noise.

\begin{table}[h]
    \caption{Core functionalities of the \textit{CategoricalClassification} framework.}
    \label{tab:core-func}
    \begin{tabularx}{\textwidth}{lX}
        \toprule
        \textbf{Functionality}      & \textbf{Description} \\
        \midrule
        Feature Generation          & We can generate features with specific value domains \& distributions (e.g. long-tail distribution), or use default settings to generate features of set feature cardinalities following normal distributions. \\[30pt]
        Target Vector Generation    & We can assign class values with built in functions (via clustering, linear, or nonlinear combinations), or define a custom decision function. \\[20pt]
        Correlations \& Combinations & The framework generates numeric features, enabling easy and quick vector computations of feature combinations. Feature correlations are computed by vector rotations. \\[30pt]
        Data Augmentation           & We can easily introduce a desired level and type of noise, or simulate missing data. \\[20pt]
        Modularity \& Customization & The framework is designed with modularity in mind. All functionalities that define relationships present in the generated dataset also support user defined functions (e.g. a specific feature combination, or class relationship).\\
        \bottomrule
    \end{tabularx}
\end{table}

\begin{figure}[ht!]
    \centering
    \subfigure{}{\includegraphics[width=.47\linewidth]{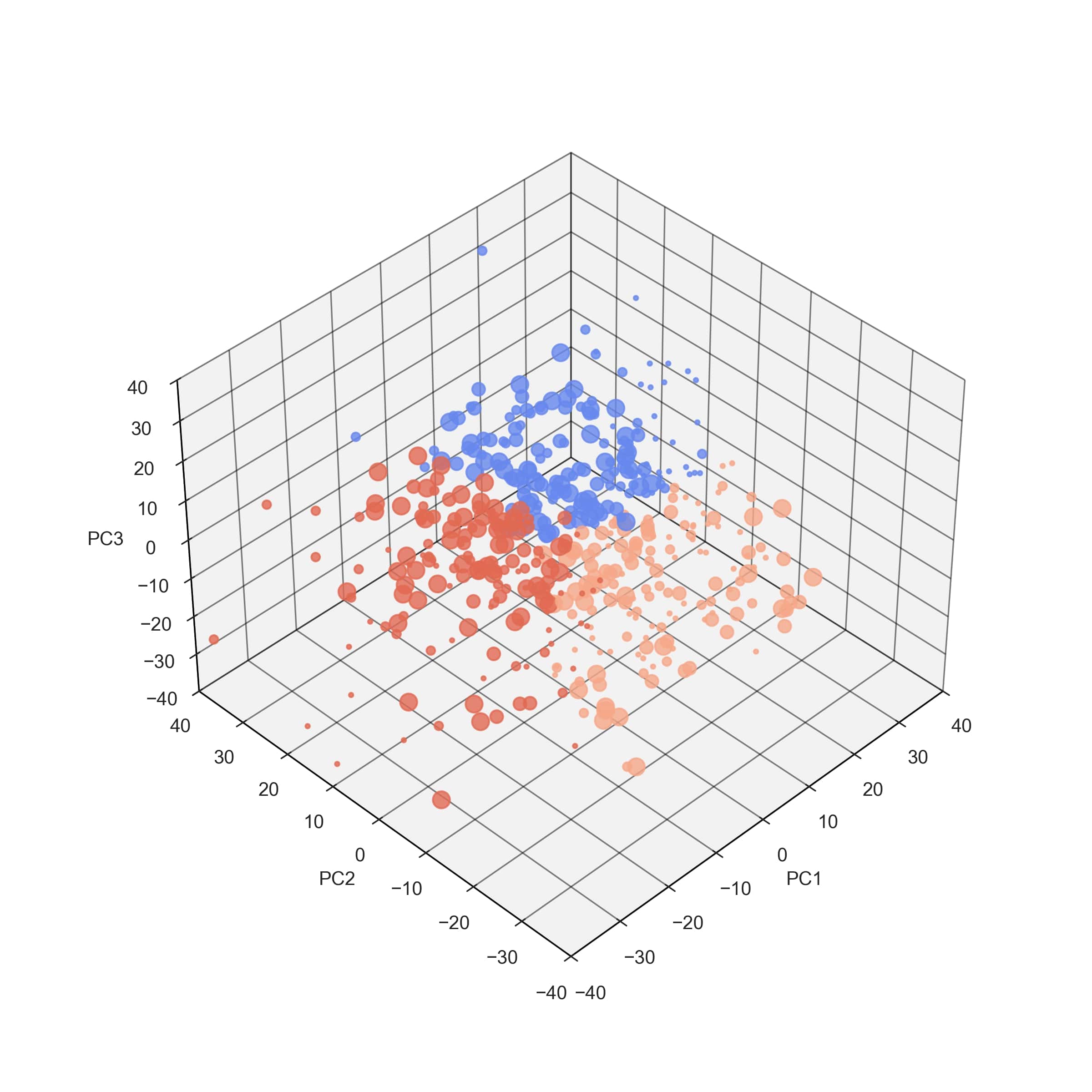} }
    \subfigure{}{\includegraphics[width=.47\linewidth]{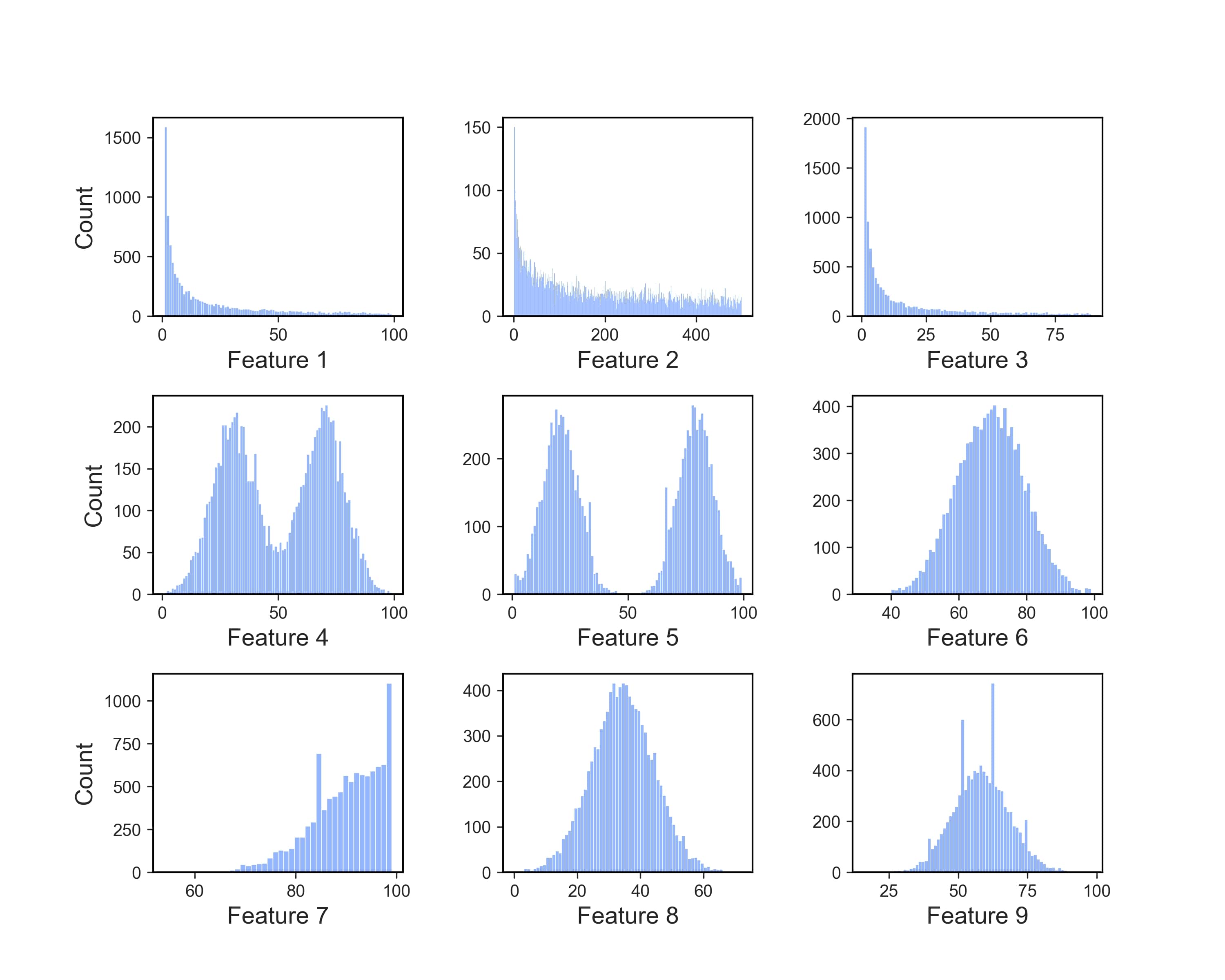} }
    \caption{\textbf{PCA plot (left) and feature densities (right) of a synthetic dataset with 9 features and 10000 samples generated with our framework.} To demonstrate its capabilities, we generated two target vectors, one via clustering (color), and one using a custom defined decision function (size), seen in the PCA plot. Using the structure parameter, we created features with differing distributions, including the commonly seen long-tail, bimodal, and normal distributions.}
    \label{fig:data}
\end{figure}

\noindent
The \textit{CategoricalClassification} framework generates datasets comprised of integer arrays which represent various categorical values, from encoded categories, hashes, to counts. All generation processes are reproducible through the use of random seeds. For simpler datasets, a single function call can generate a useful, synthetic dataset, while the functionalities described in table \ref{tab:core-func} streamline the generation of more complex datasets with specific attributes to fit experimental needs. These can be specific feature value sets and distributions, specific feature-class relationships, or user defined feature combinations. To demonstrate its capabilities, the example dataset seen in figure \ref{fig:data} was generated, featuring various types of distributions and feature cardinalities. Adding combinations of features or correlated features, is a simple matter of calling the appropriate function and specifying the desired (column) indices.

To ensure familiarity, the API structure of our framework follows common principles seen in libraries like \textit{NumPy} or \textit{SciPy}. The framework itself is available as an integrated module of the open-source tool \textit{Outrank} (\url{https://github.com/outbrain/outrank}), or as a standalone tool \textit{catclass} (\url{https://github.com/98MM/msc_cc}), both installable using \textit{pip}.

\section{Applications of custom synthetic datasets}
In this section we consider three applications that showcase how custom generated synthetic datasets can give novel insights into methods commonly used in recommender systems.

\subsection{Use Case 1 -- Benchmarking Probabilistic Counting Algorithms}
\label{sec:bench-pc}

\begin{figure}[ht!]
    \centering
    \subfigure{}{\includegraphics[width=.45\linewidth]{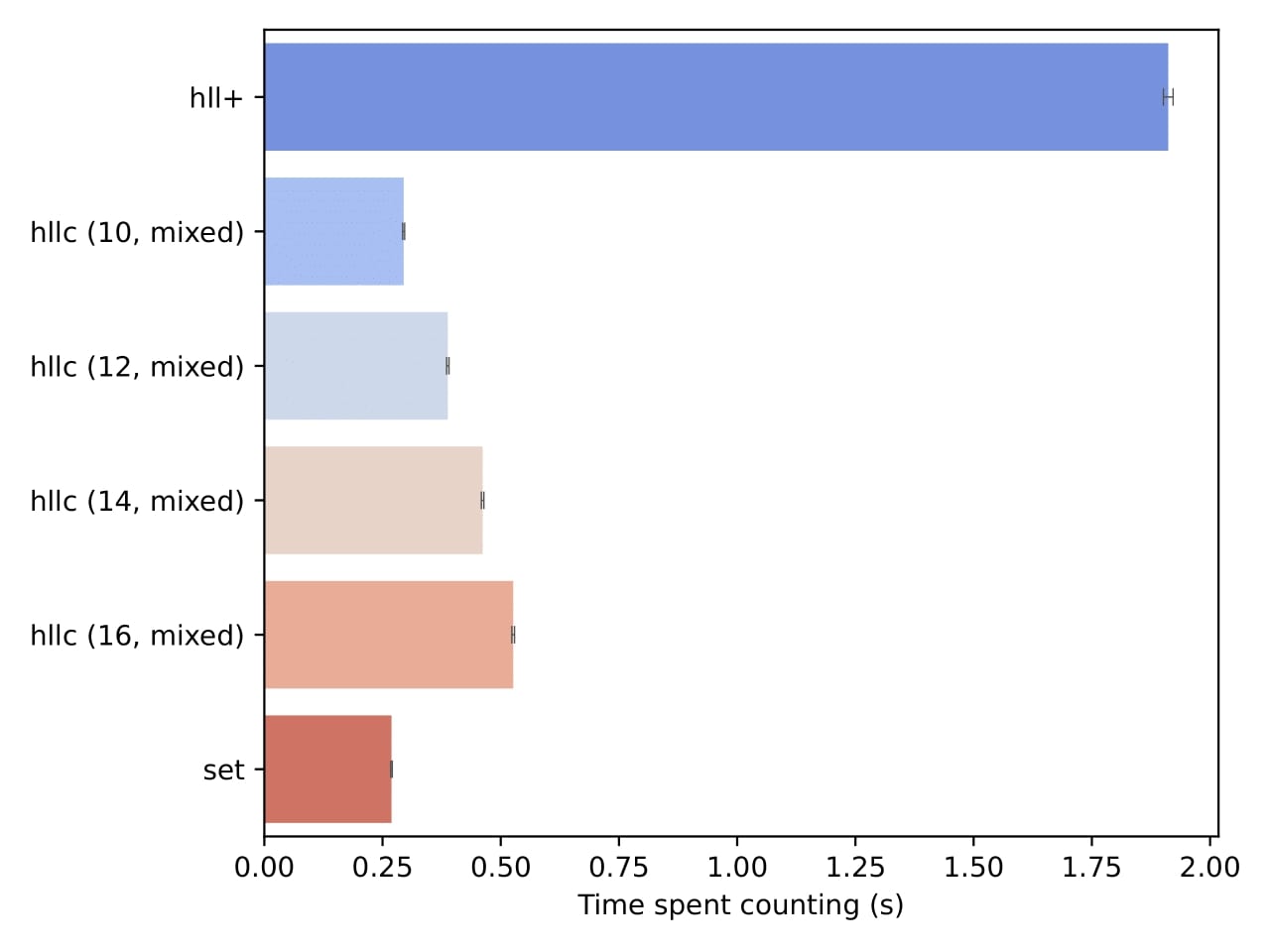} }
    \subfigure{}{\includegraphics[width=.45\linewidth]{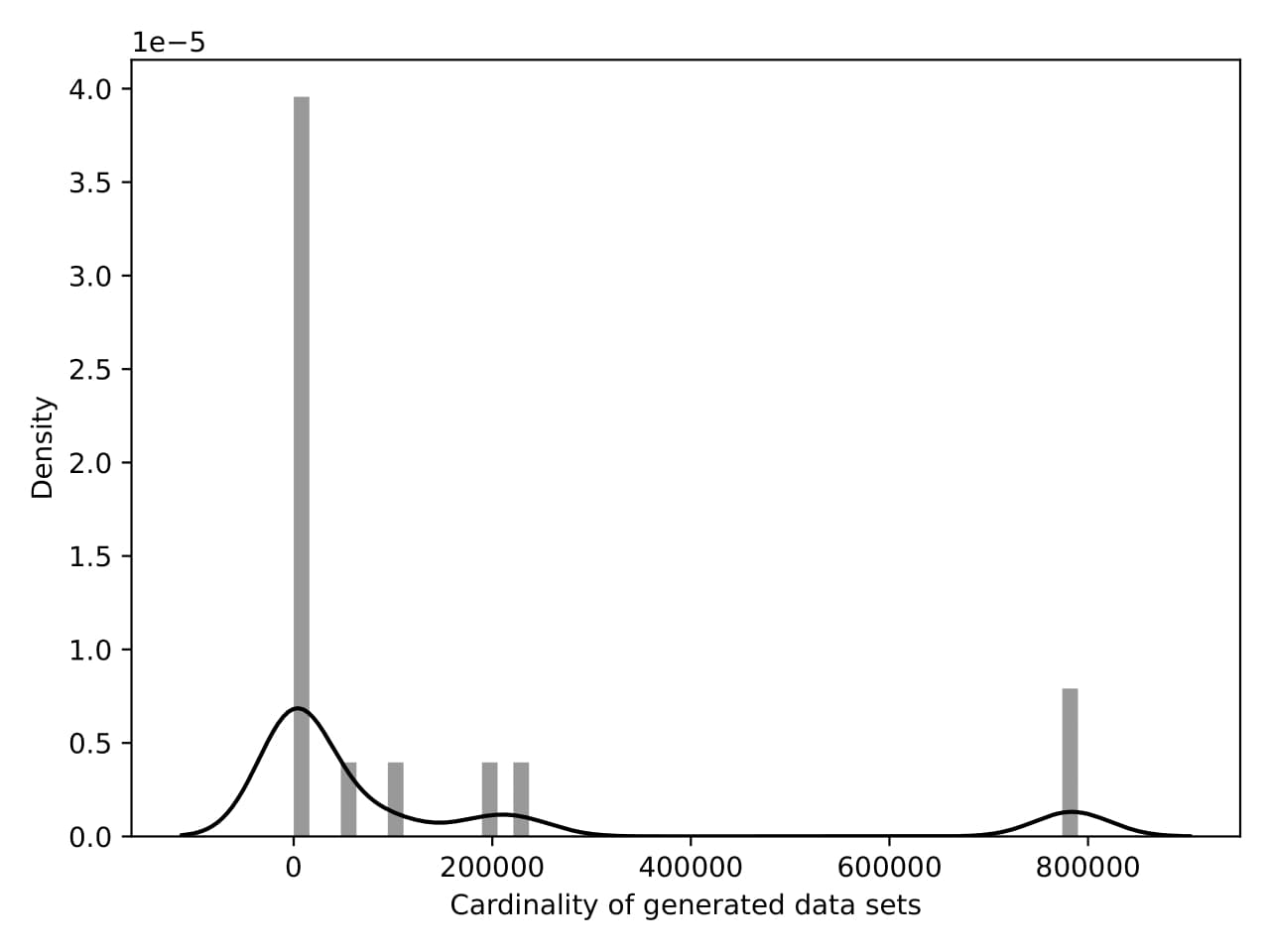} }
    \caption{\textbf{Distributions of computation times over more than 2k synthetic datasets comprised of 20 features and 1m rows.} All algorithms ensure error rate smaller than 0.005 (the set is exact). Small-enough hllc -- hyperloglog with caching performs with minimal error and similar times to set itself -- this is due to the fact that in most cases, it remains deterministic for most of the datasets.}
    \label{fig:prob-count}
\end{figure}

Categorical data streams are ubiquitous in modern recommender systems. Features include different categories, identifiers, or aggregates of numeric features and similar counts. When monitoring online streams, effectively counting the number of unique items in the stream becomes a challenging computational problem. Exact counting methods (such as hashing) adhere to substantial memory overhead that seldom scales in practice. To remedy this shortcoming, probabilistic counting algorithms were introduced. In particular, we're interested in the HyperLogLog family -- algorithms aimed at estimation of the number of unique items (cardinality) in a data batch (part of a stream). We observed that probabilistic counters, albeit much more memory efficient, introduce some noise in terms of precision -- which is expected. However, the issue is that common algorithms don't discriminate between low- and high-cardinality items. This in practice means that a low-cardinality feature, where estimate is of high impact (e.g., count of days in a week) is occasionally miscounted, resulting in errors that carry bigger impact than e.g., counting unique users on a site. To remedy this shortcoming, we introduce a caching mechanism to arbitrary HyperLogLog-like algorithm, where, to a certain degree, the algorithm remains deterministic, and only switches to probabilistic mode of operation if its memory requirements exceed the constrained (user-specified) value. See Figure \ref{fig:prob-count} for a visualization of these results.

\subsection{Use Case 2 -- Detecting Algorithmic Bias}
\label{sec:bias}

Datasets with complex feature interactions present significant challenges for machine learning models. As the complexity of these interactions increases, so does the difficulty in accurately modelling and capturing these relationships. Moreover, complex feature interactions can lead to overfitting or underfitting, resulting in poor generalizations to unseen data, or the inability to capture intricate relationships in the data. Traditional linear models, such as the logistic regression are efficient, but unfortunately often fail to capture complex feature interactions. Advanced models, such as the DeepFM model which combines factorization machines and deep neural networks \cite{shen2017deepctr, guo2017deepfm}, are better equipped to handle such complexities. As such we are interested in how Logistic Regression and DeepFM models compare when presented with datasets with increasingly complex feature interactions. To systemically evaluate bias and performance, we generate an initial synthetic dataset with 4 relevant and 750 irrelevant features, 10k samples and a nonlinear class relation. We introduce 20\% of categorical noise and create various feature interactions based on pairwise combinations of relevant features, which are subsequently removed from our generated dataset. We then iteratively remove the resulting combination features, perform minimal hyperparameter tuning and evaluate DeepFM and logistic regression performance over one epoch. When multiple types of combinations are present in the generated dataset, we observed an increase in both the AUC and accuracy scores for both models.

\begin{figure}[ht!]
    \centering
    \subfigure{}{\includegraphics[width=.47\linewidth]{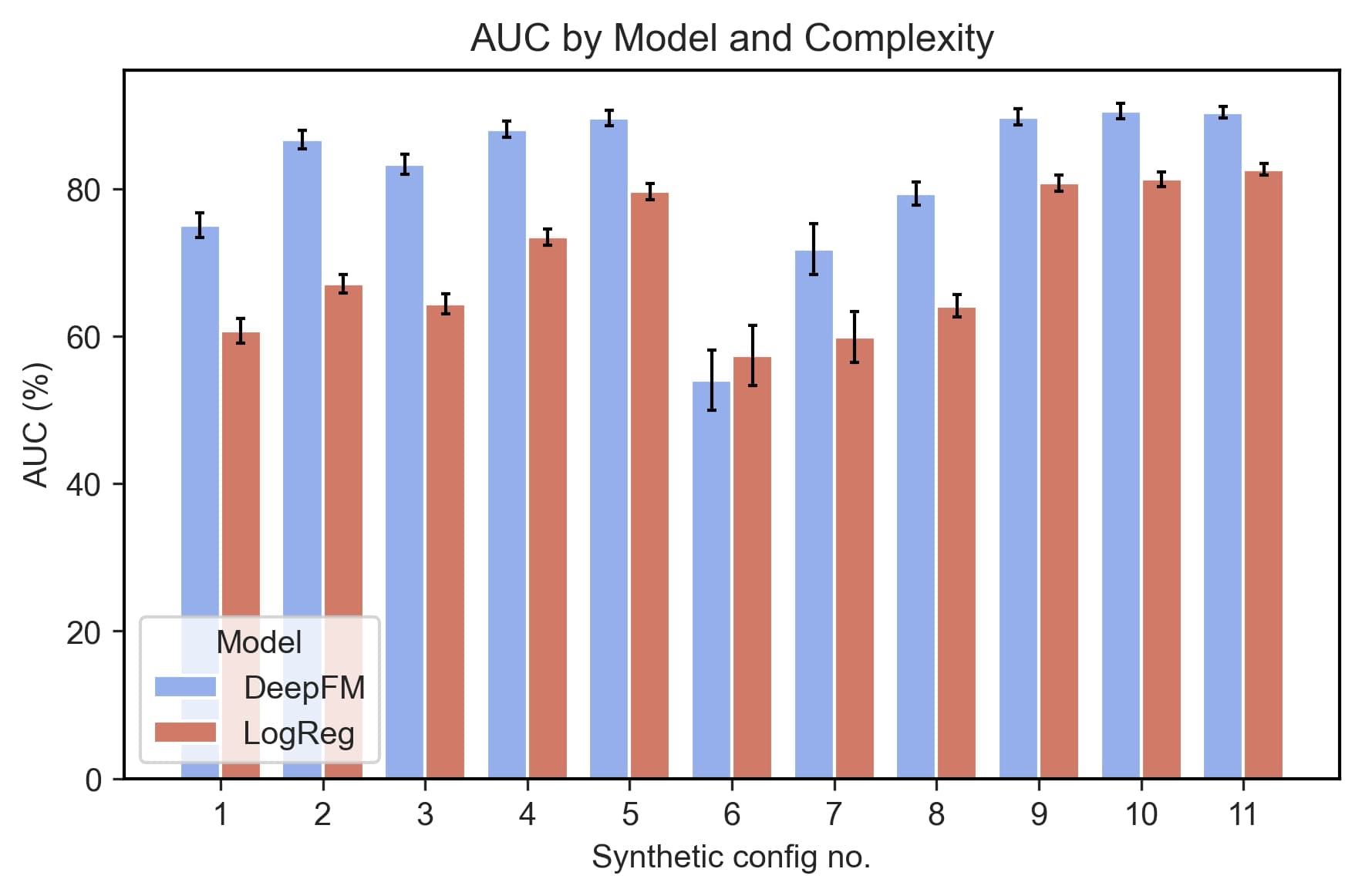} }
    \subfigure{}{\includegraphics[width=.47\linewidth]{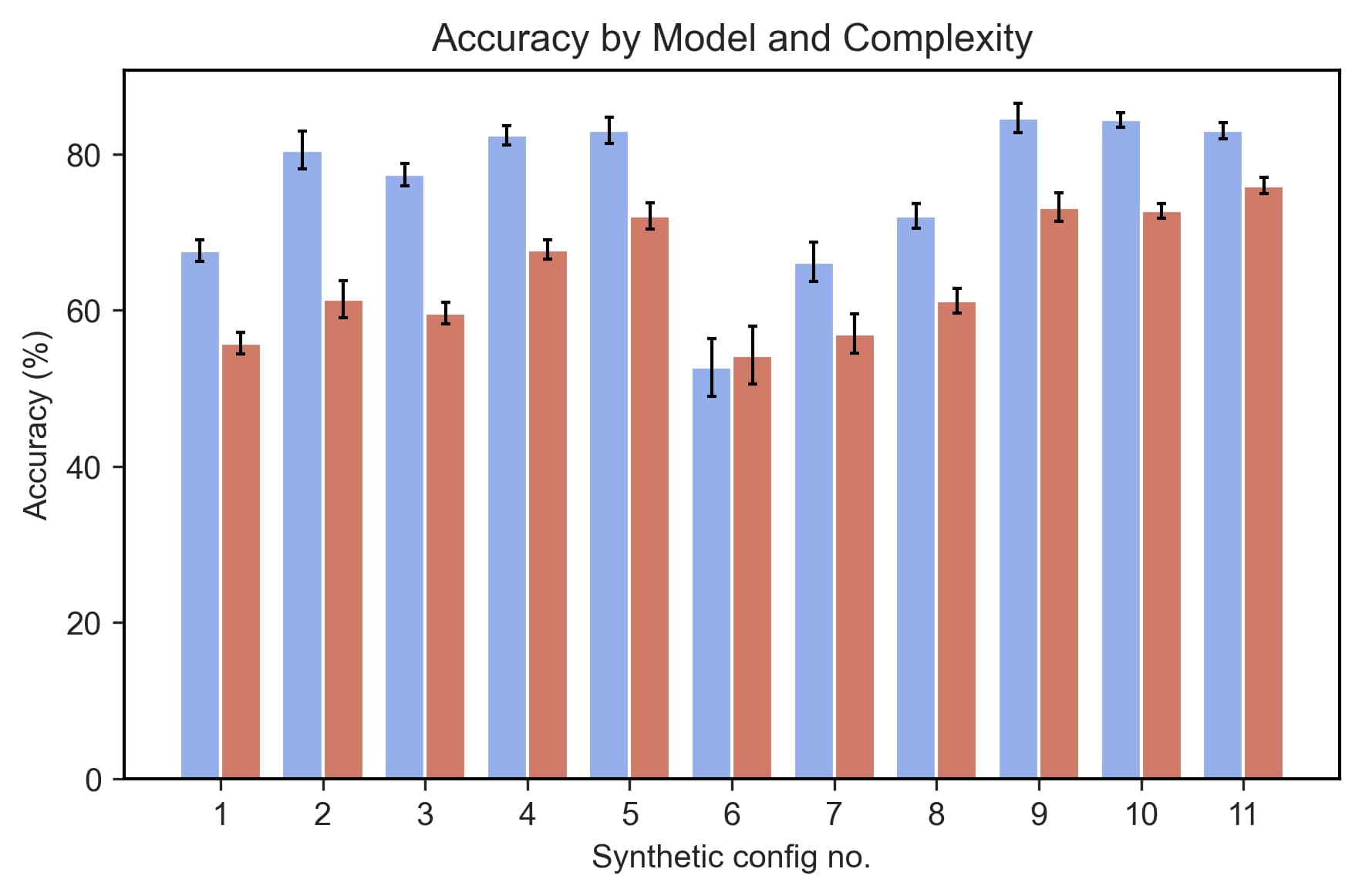} }
    \caption{\textbf{AUC (left) and accuracy (right) scores of DeepFM and logistic regression after one epoch.} Synthetic dataset configs are sets of pairwise relevant feature combinations -- 1: AND, 2: OR, 3: XOR, 4: AND, OR, 5: AND, OR, XOR, 6: sum of squares, 7: square of sums, 8: both square combinations, 9: AND, OR, XOR, sum of squares, 10: AND, OR, XOR, square of sums, 11: all feature combinations present.}
    \label{fig:bias}
\end{figure}

As seen in figure \ref{fig:bias} DeepFM significantly outperformed logistic regression in all but one case -- when presented with features created with a sum of squares combination. %See Figure \ref{fig:bias} for a visualization of these results. (redundant, as the first sentence already refrences figure)

\subsection{Use Case 3 -- Simulating AutoML search}
\label{sec:automl}

Automated Machine Learning (AutoML) represents a paradigm shift in the way machine learning models are developed and deployed. AutoML aims to decrease the time and effort required to develop robust machine learning solutions by automating data pre-processing, feature ranking and engineering, model selection, hyperparameter tuning and model evaluation.

Building effective recommender systems involves managing complex feature interactions in a wide feature space and often hinges on the fine-tuning of its underlying models and algorithms. One of the most important AutoML operations in recommender systems is feature selection. More precisely, the question of finding the smallest subset of features that will, given an ML algorithm, deliver optimal, or close to optimal performance. In this way one optimizes the size of a model which is especially valuable for deployments on a large scale and when we are dealing with memory limitations.

A commonly used strategy is to iteratively grow the feature set by adding the next best feature that is not already in the set. In the first step we find the best feature $f_1$ by finding the model that gives best predictions with a single feature. After that, we find the best model predicting with $f_1$ and another feature, called $f_2$, etc. We repeat the process until we have a feature set of the desired size or until we no longer see a performance boost. In this use case we use the aforementioned open source package called Outrank. Outrank implements the functionality for the iterative feature set growth -- given an existing subset of features from our dataset, provide the ranking of the remaining features by evaluating the performance of all models containing the existing features and one of the remaining ones. For this we use the so-called \textit{surrogate-SGD-prior} heuristic of Outrank, which, under the hood, implements Scikit-learn's SGDClassifier and runs 4-fold cross-validation score with a negative log loss.

\begin{figure}[ht!]
    \centering
    \subfigure{1}{\includegraphics[width=.31\textwidth]{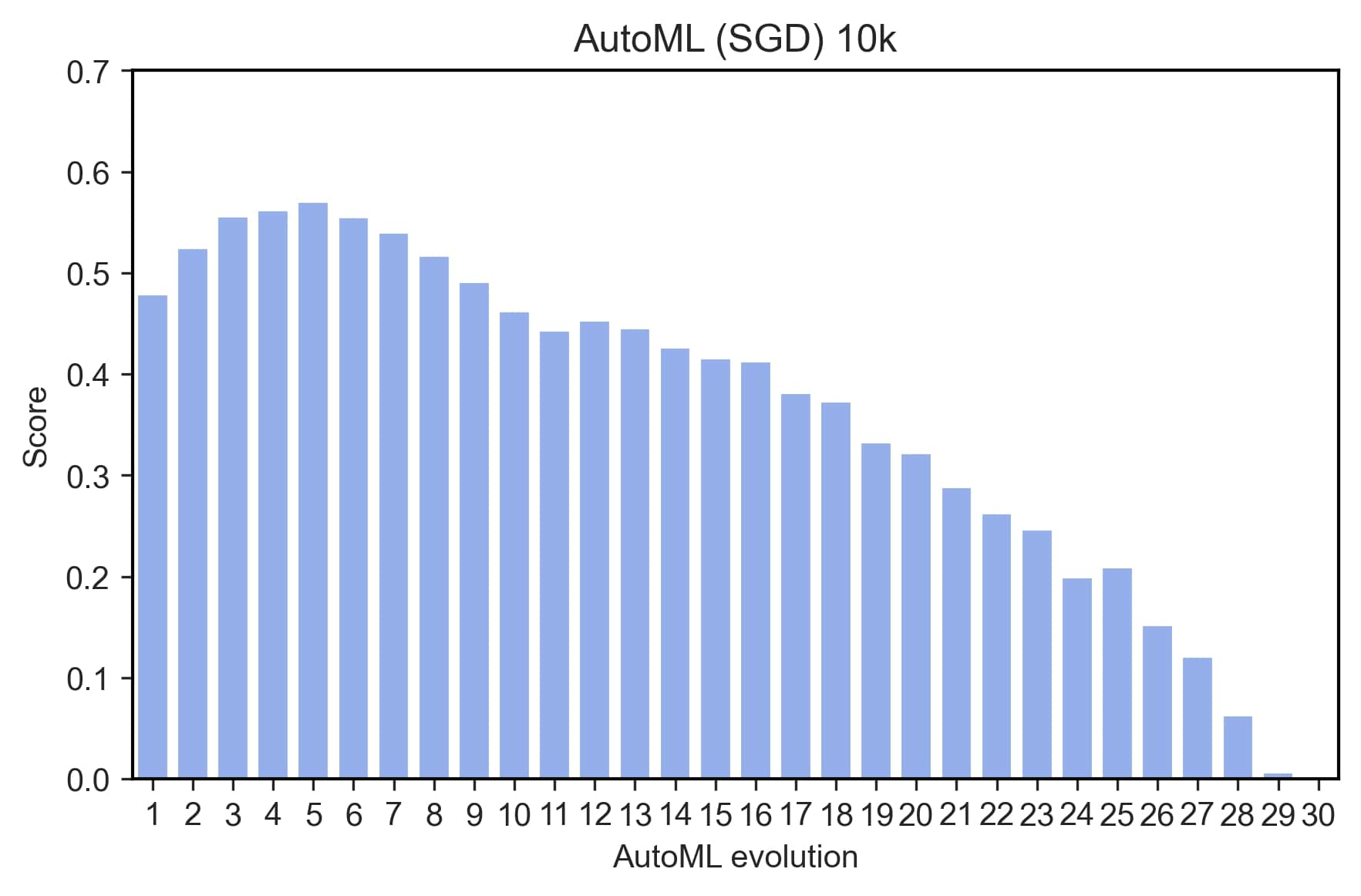} }
    \subfigure{2}{\includegraphics[width=.31\textwidth]{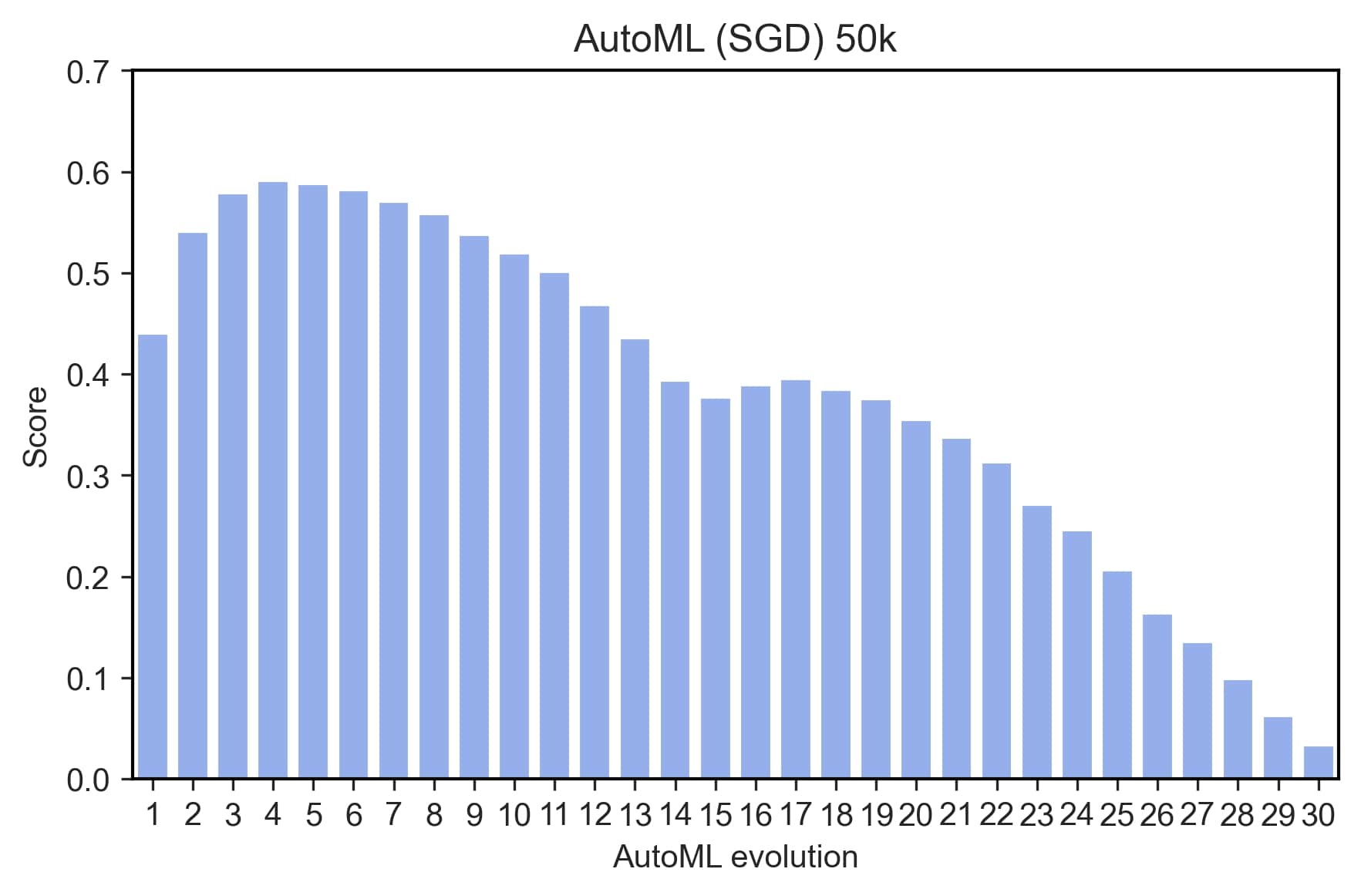} }
    \subfigure{3}{\includegraphics[width=.31\textwidth]{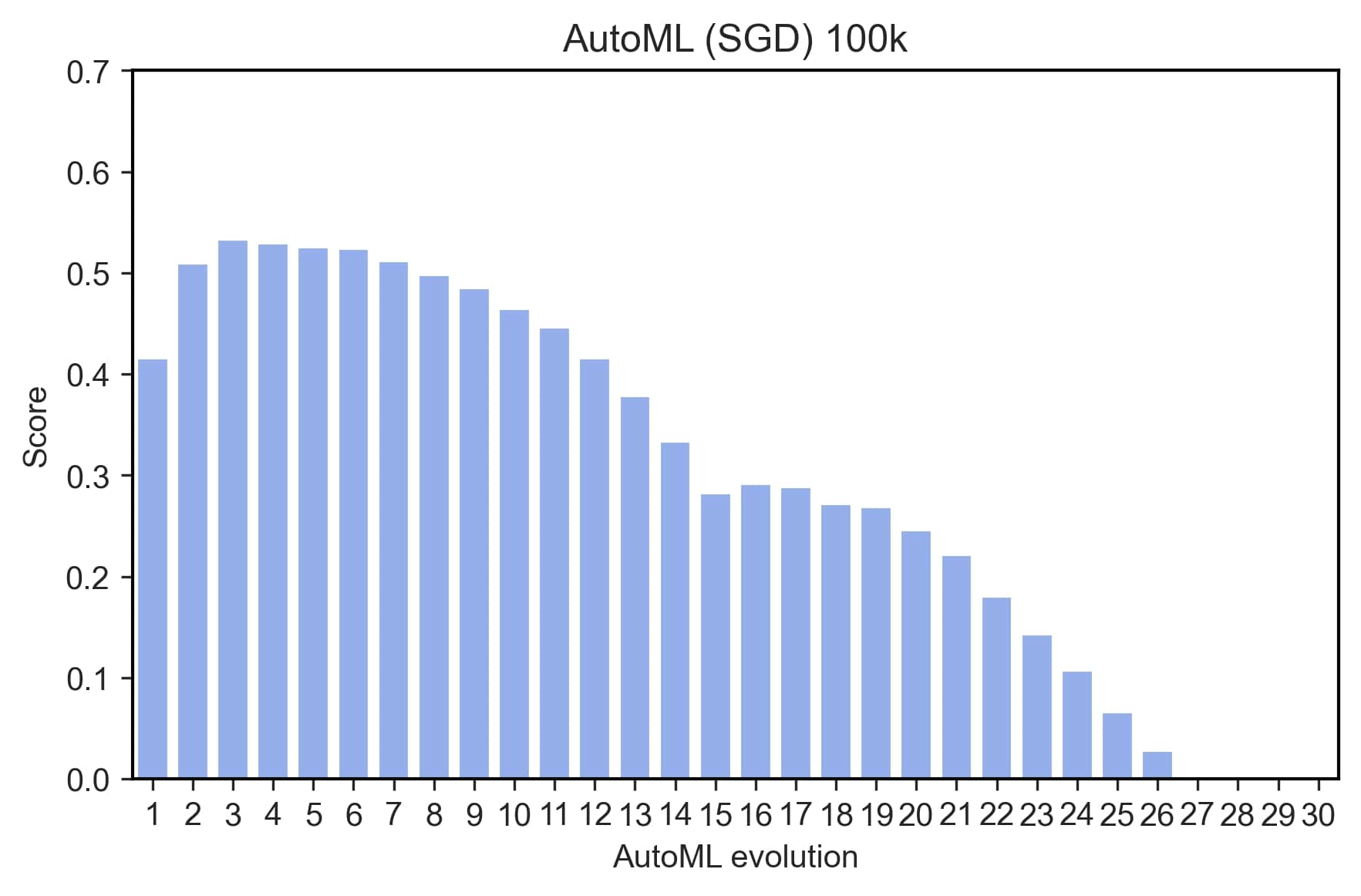} }
    \caption{\textbf{Negative Log Loss for AutoML evolution with different dataset sizes.} \\Evolution set 1: \{OR3, OR4, AND2, XOR0, AND3, IRR50, IRR75, IRR27, AND4, IRR15, SUM\_SQUARES2...\} \\Evolution set 2: \{OR1, AND4, OR4, OR3, AND2, OR0, AND0, AND5, OR2, OR5, XOR5, IRR69, AND1, AND3...\} \\Evolution set 3: \{OR5, OR1, AND0, AND5, OR3, OR2, AND4, AND3, OR4, OR0, AND1, AND2, IRR3, IRR41...\}}
    \label{fig:automloutrank}
\end{figure}

For simulations of such AutoML searches, we generate three initial datasets with 4 relevant and 900 irrelevant features, and a nonlinear class relation, with 10k, 50k, and 100k samples. As in section \ref{sec:bias}, we introduce 20\% of categorical noise and create various feature interactions based on pairwise combinations of relevant features which are subsequently removed from our generated dataset. Figure \ref{fig:automloutrank} shows the evolution of scores of the models, when adding new features.

\begin{figure}[ht!]
    \centering
    \subfigure{}{\includegraphics[width=.32\textwidth]{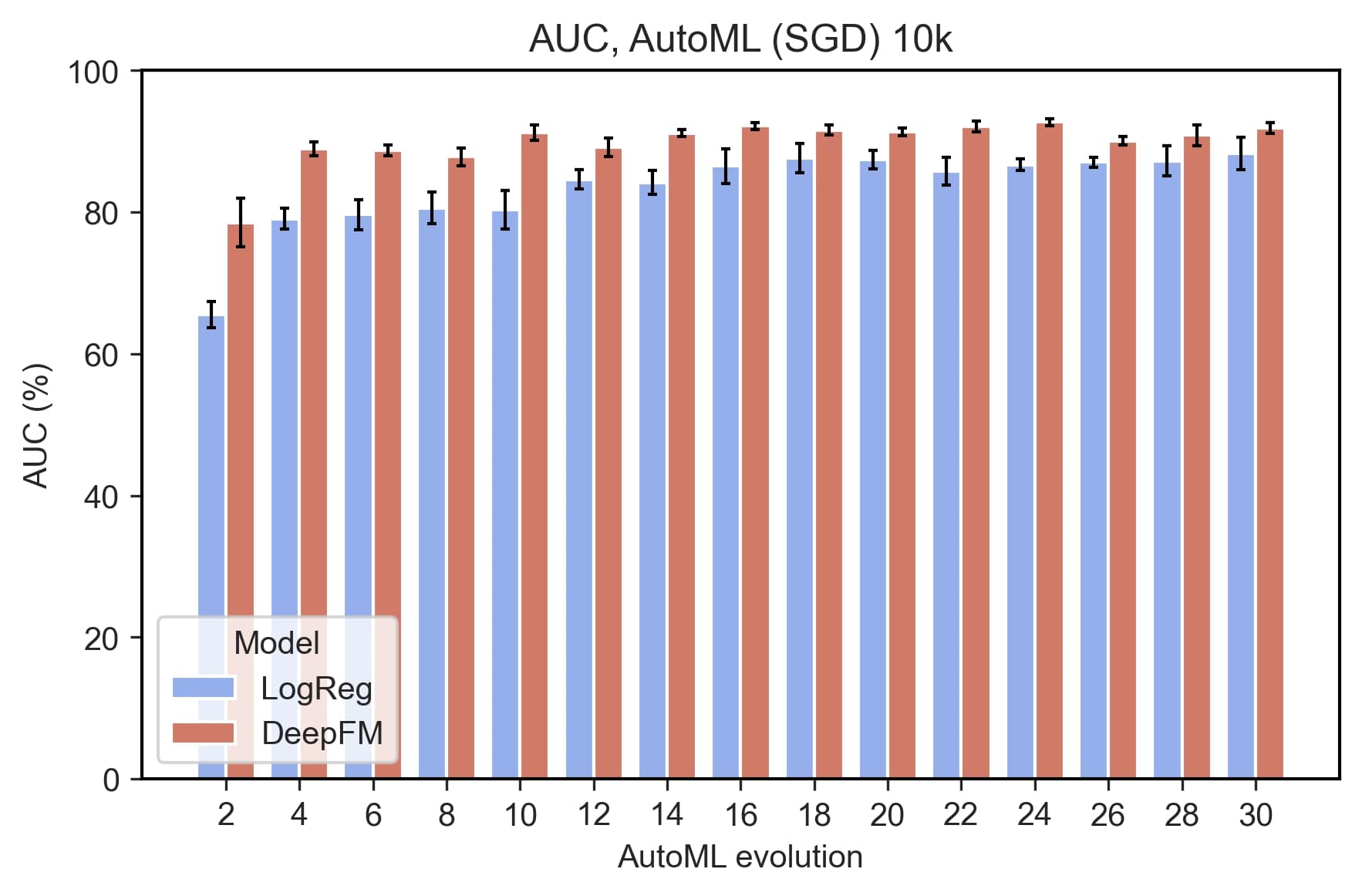} }
    \subfigure{}{\includegraphics[width=.32\textwidth]{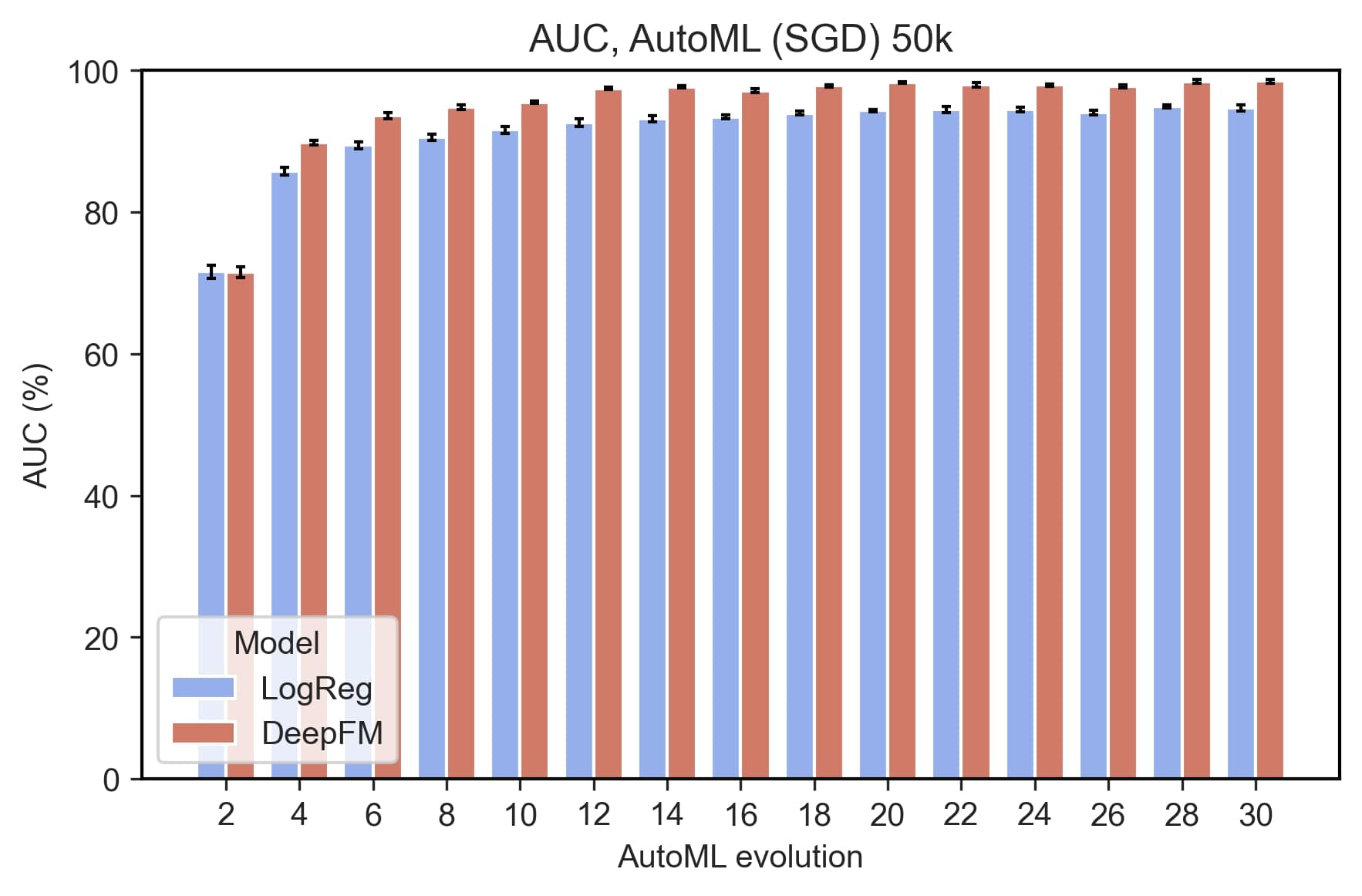} }
    \subfigure{}{\includegraphics[width=.32\textwidth]{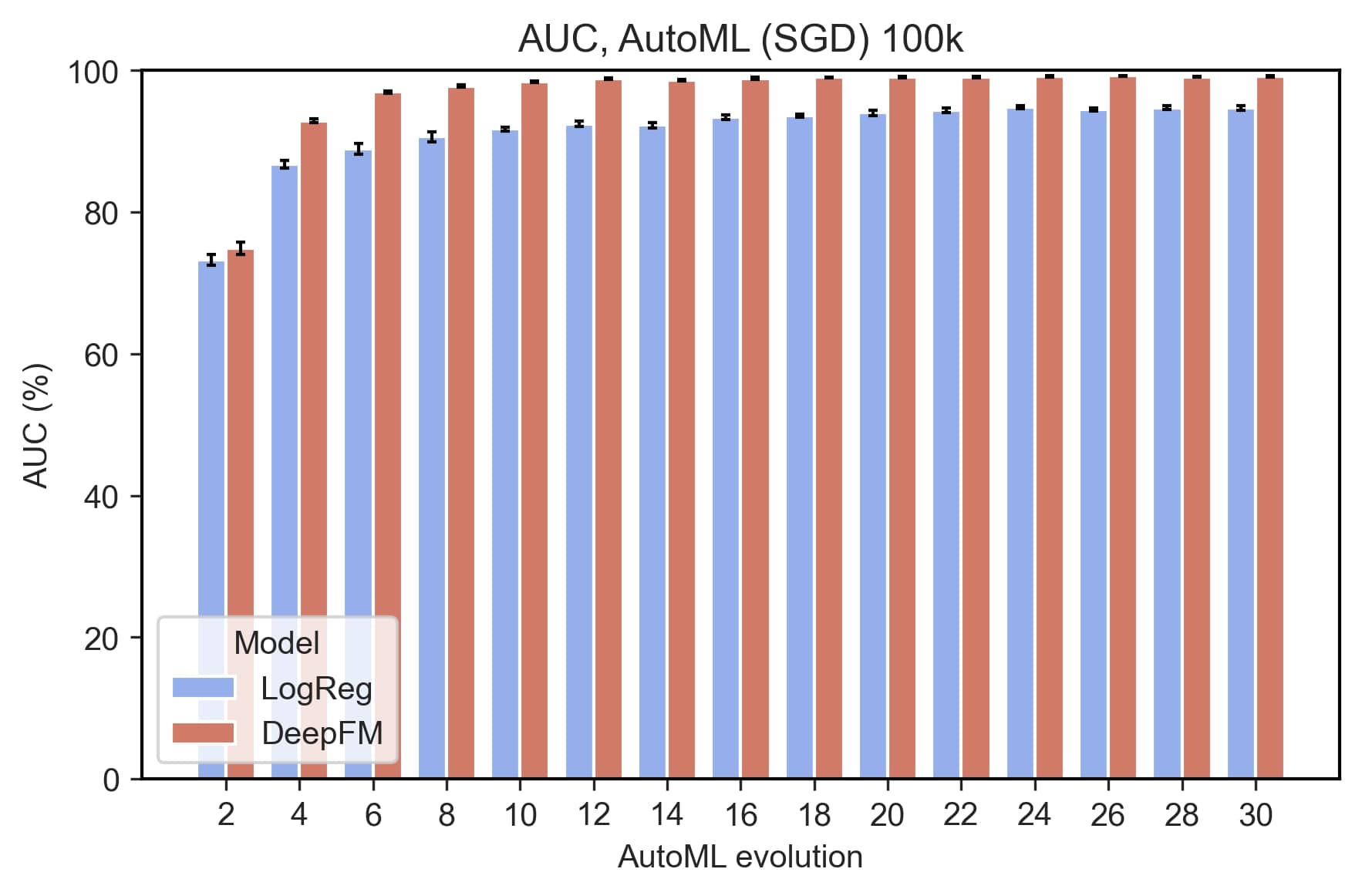} }
    \subfigure{}{\includegraphics[width=.32\textwidth]{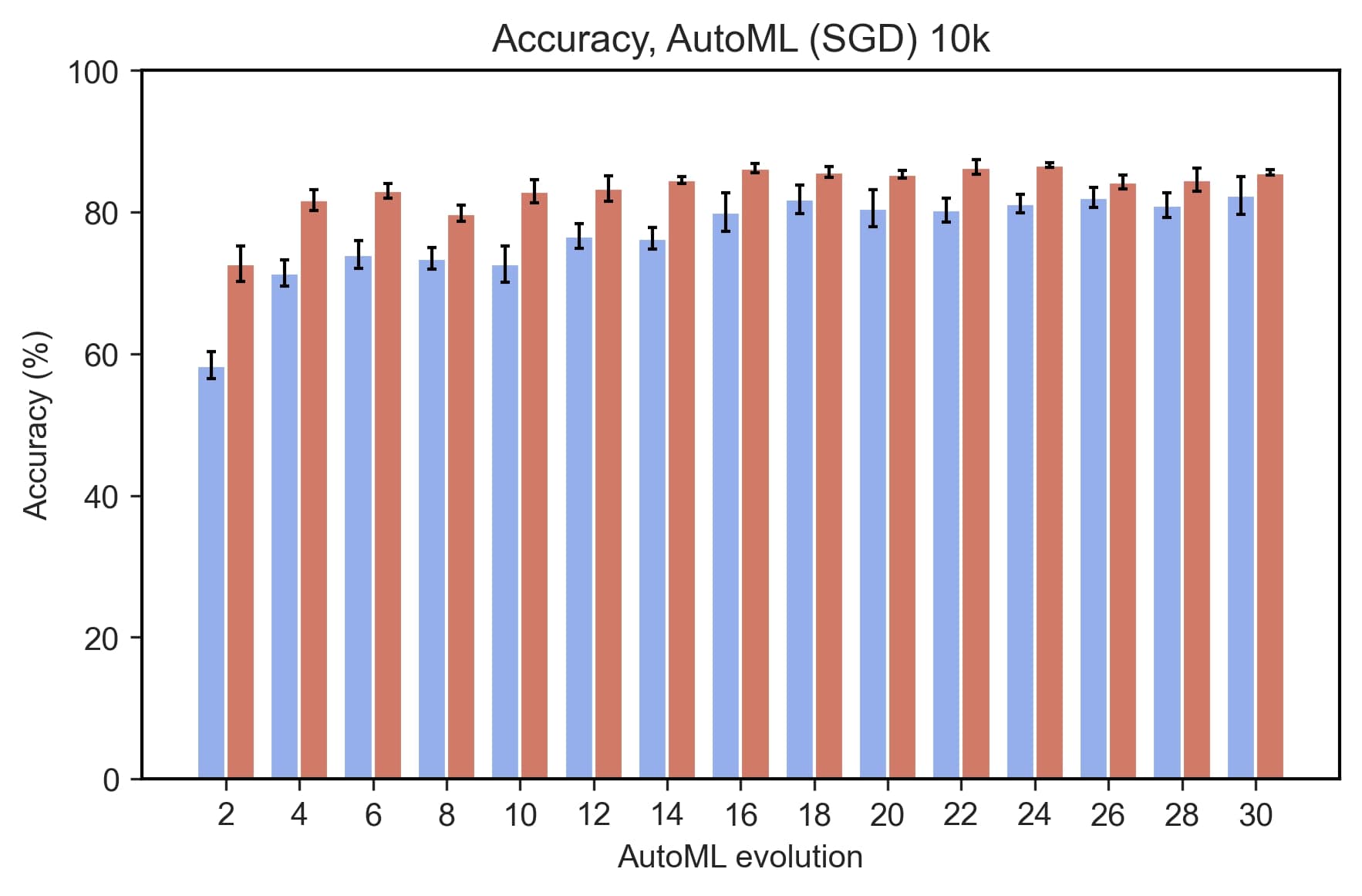} }
    \subfigure{}{\includegraphics[width=.32\textwidth]{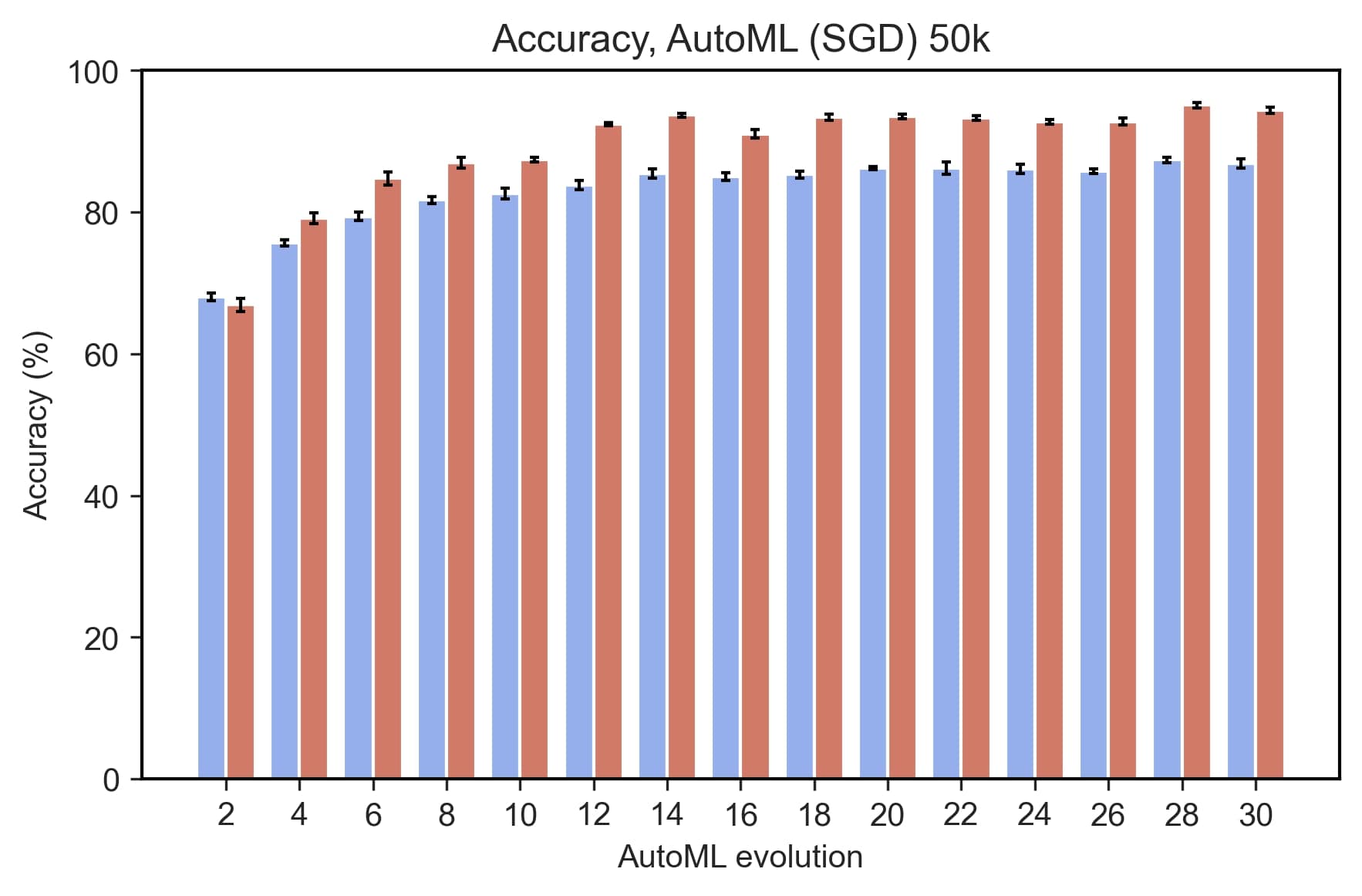} }
    \subfigure{}{\includegraphics[width=.32\textwidth]{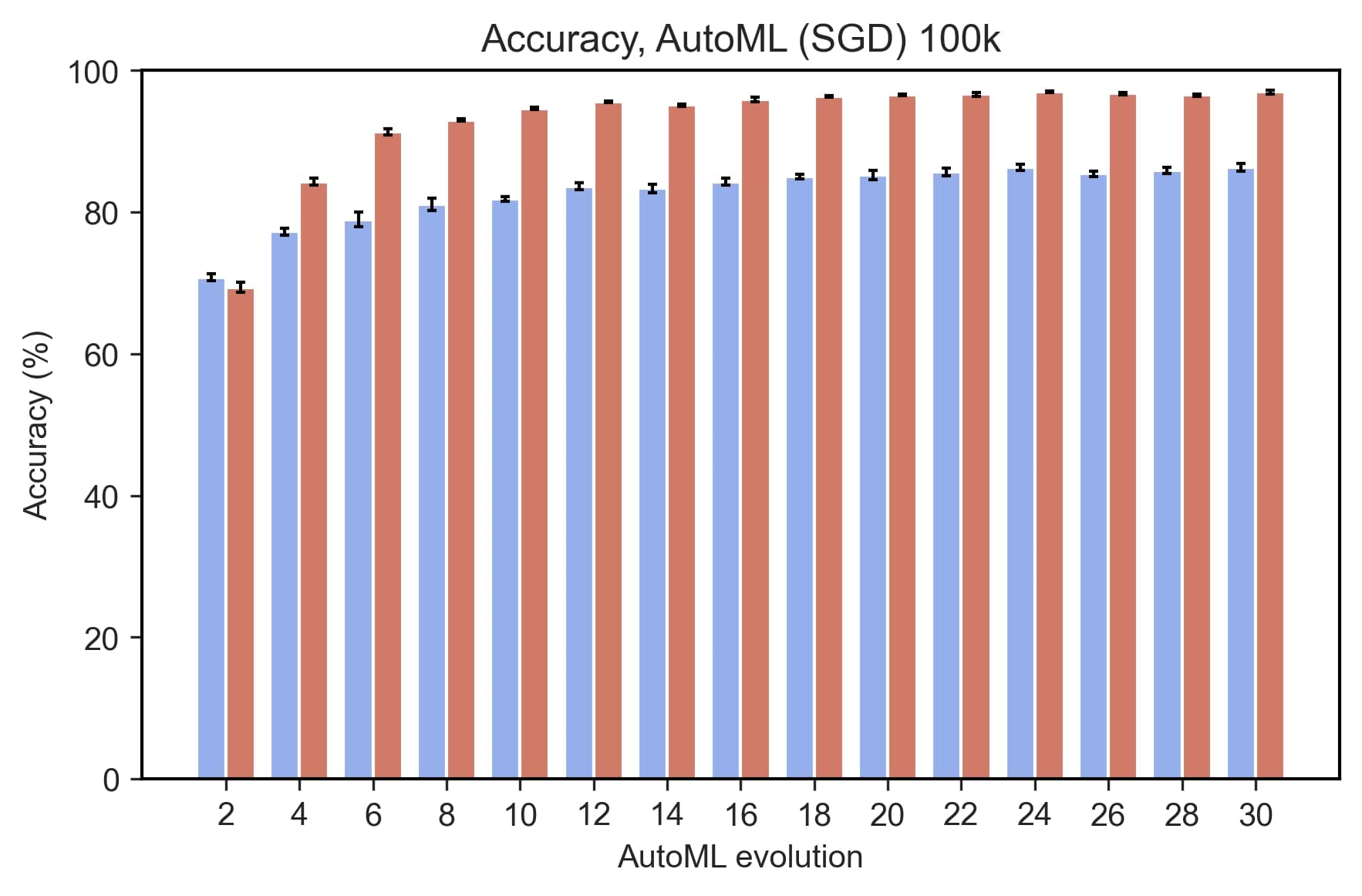} }
    \caption{\textbf{Accuracy scores and AUC of DeepFM and logistic regression for features from AutoML and different datasets.}}
    \label{fig:automlfinal}
\end{figure}

Important to note here is that due to the complexity of the search, Outrank has no hyperparameter optimisation capability for the models that provide feature rankings. This explains why the results in figure \ref{fig:automloutrank} are falling very fast with the increase of feature complexity. The performance of the models with hyperparameter tuning has thus been evaluated similarly as in the section \ref{sec:bias} (see Figure \ref{fig:automlfinal}). The positive trend in the AutoML evolution of Outrank (figure \ref{fig:automloutrank}) seems to imply a positive trend in the performance of the corresponding models trained with hyperparameter tuning. This, on one hand, is useful information for optimal feature selection. On the other hand, it shows that AutoML for feature selection without hyperparameter optimisation can be misleading.

\section{Discussion and Conclusions}

The framework presented in this paper offers a robust and versatile tool for creating synthetic datasets for testing and evaluation of real-world recommender systems. By allowing precise control over dataset attributes, researchers can design experiments that isolate specific aspects of algorithm performance, especially when investigating scenarios that are not commonly encountered in real-life data. Our use cases demonstrate the framework's practical application.

We showcased the usefulness of our synthetic data generation framework on several real-world scenarios. In the first one, we precisely evaluated probabilistic counting algorithms. By testing the algorithms on more than 2k synthetic dataset we were able to highlight a key limitation -- the probabilistic counting algorithms' inability to discriminate between low- and high-cardinality items. By introducing a caching mechanism, we enhanced an arbitrary HyperLogLog-like alogrithm's precision for low-cardinality features.

In the second example, we used our framework's ability to generate complex feature interactions to systematically evaluate the performance of logistic regression and DeepFM models on datasets with varying levels and amounts of feature interaction complexity. The results demonstrated DeepFM's superior ability to handle complex interactions, significantly outperforming logistic regression.

Finally, we simulated AutoML search and evaluated its ability to identify relevant features. Through these use cases we demonstrated our framework's efficacy in generating challenging datasets to simulate real-world scenarios in a controlled environment.

The three use cases demonstrate our framework's capabilities at various stages in the recommender systems' pipeline. Evaluating probabilistic counting algorithms may decrease memory overhead when determining feature cardinality. The framework's ability to simulate properties of real-life data in a controlled manner enables us to generate data tailored to a specific problem, enabling further insight into model behavior (i.e. determining algorithm bias). By controlling the generation process we also control feature relevance, enabling us to test key functionalities of AutoML systems integrated within many recommender systems' pipelines.

Despite its strengths, there are areas for future improvement. By integrating advanced generative models such as GANs or variational autoencoders we could further enrich the diversity and realism of the synthetic datasets. Additionally, expanding the framework to support other types of machine learning tasks, such as regression, could broaden its applicability and impact. Nevertheless, by enabling controlled, repeatable experiments, our framework provides researches and practitioners with a powerful tool to advance the field, ultimately leading to more effective and reliable recommender systems.

\bibliography{bibliography}
\end{document}